\documentstyle[12pt]{article}
\newcommand{\bff}[1]{{\mbox{\boldmath $#1$}}}
\begin{document}
\setlength{\baselineskip}{0.7cm}
\vspace{2.0cm}
\centerline{\Large \bf Superdeformations in Relativistic and} 
\vspace{0.5cm}
\centerline{\Large \bf Non-Relativistic Mean Field Theories}
\vspace{1.0cm}
\centerline{\Large A.\ V.\ Afanasjev \footnote{E-mail: 
anatoli.afanasjev@Physik.TU-Muenchen.DE} 
\footnote{Alexander von Humboldt
fellow. On leave from Laboratory of Radiation Physics,
Institute of Solid State Physics, University of Latvia,
LV 2169, Salaspils, Miera str 31, Latvia}
and P.\ Ring}
\vspace{0.5cm}
\centerline{\large Physik-Department der Technischen Universit{\"a}t
M{\"u}nchen,}
\centerline{\large D-85747 Garching, Germany}
\vspace{1.5cm}
\centerline{\Large \bf Abstract}
The applications of the extensions of relativistic mean field 
(RMF) theory to the rotating frame, such as cranked relativistic 
mean field (CRMF) theory and cranked relativistic Hartree-Bogoliubov 
(CRHB) theory, for the description of superdeformed
bands in the $A\sim 60$, $140-150$ and 190 mass regions are 
overviewed and compared briefly with the results obtained in
non-relativistic mean field theories.

\vspace{1.5cm}
\centerline{PACS: 21.60.-n, 21.60.Cs, 27.50.+e, 27.60.+j, 27.70+q, 27.80.+w}
\input epsf

\newpage
\setlength{\baselineskip}{0.5cm}
\section{Introduction}

\begin{figure}
\epsfxsize 16.0cm
\epsfbox{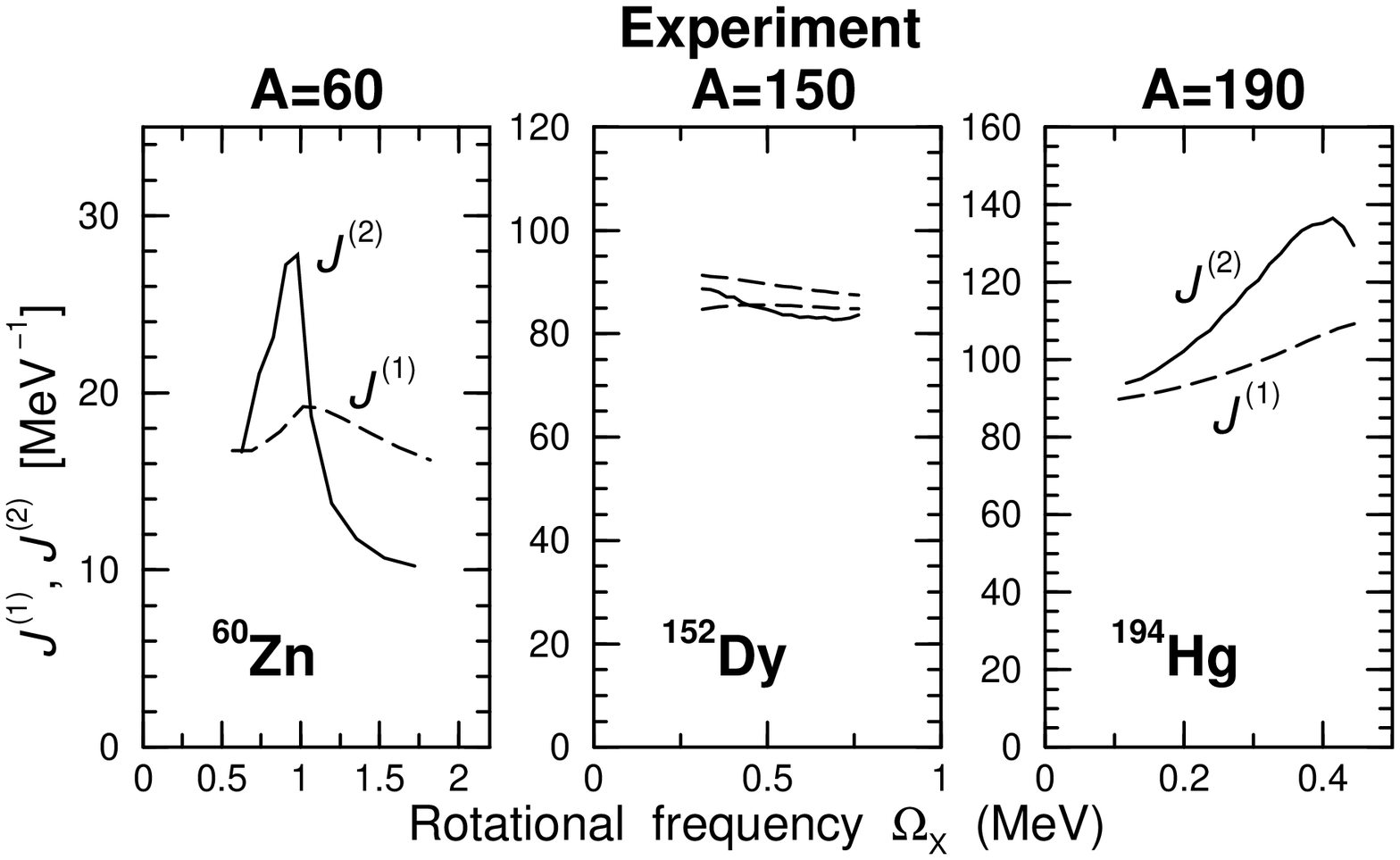}
\caption{Dynamic ($J^{(2)}$) and kinematic ($J^{(1)}$) moments 
of inertia of representative SD bands observed in the $A\sim 60$, 
150 and 190 mass regions. SD bands in $^{60}$Zn and $^{194}$Hg 
are linked to the low-spin level scheme. On the contrary, no
SD band in $A\sim 140-150$ mass region is linked to the 
low-spin level scheme. Thus the $J^{(1)}$ values for the 
SD band in $^{152}$Dy are shown under two different spin 
assignments: the lowest transition in SD band with energy
602.4 keV corresponds to the spin changes of $26^+\rightarrow 24^+$ 
(lowest curve) and of $28^+\rightarrow 26^+$ (highest curve).
One should note that the analysis based both on the effective alignment 
approach and the systematics of experimental estimations of the spins 
of the lowest states in the SD bands in this mass region favours the first 
alternative, however, the second alternative cannot be excluded,
see Ref.\ \protect\cite{ALR.98} for details.}
\label{exp-zoo}
\end{figure}

 Theoretical and experimental investigations of superdeformation 
at high spin have a long history. Starting from the prediction of 
such shapes within the macroscopic + microscopic method (frequently 
called as a cranked Nilsson-Strutinsky approach) in 70ties
\cite{SDpred1,SDpred2} and subsequent 
experimental observation of the first superdeformed rotational (SD) band
in $^{152}$Dy by P.\ Twin and collaborators in 1986 \cite{TNN.86}, 
considerable understanding of the phenomenon of superdeformation 
at high spin has been reached. 
At present, the superdeformation at high spin has been 
observed in a number of mass regions, namely, $A\sim 60-70$
\cite{Zn60,Zn68}, $A\sim 80$, 130, 150, 190 \cite{SD-sys}. As
illustrated in Fig.\ \ref{exp-zoo}, the observed SD bands show very 
different patterns for the rotational behaviour depending on
the deformation, on the strength of pairing correlations, and on the 
angular momentum content of specific SD configurations.

 The experimental progress has been accompanied by the development 
of sophisticated theoretical models which use the concepts of mean
field and the cranking model. While early detailed theoretical 
investigations of SD bands have been performed mainly within the 
macroscopic + microscopic method \cite{BRA.88,NWJ.89,SVB.90,WCNWJ.91},
the fast increase of computational power allowed the development of 
non-relativistic microscopic models based on effective forces of Skyrme
\cite{Skyrme-rot} or Gogny \cite{GDBL.94,ER.93} type and cranked
versions of the relativistic mean field theory
\cite{KR.89,KR.90,AKR.99}. Review articles concerning the applications
of these theoretical models to the SD bands can be found in 
Refs.\ \cite{AFN.90,BHN.95,D.99}.

 Relativistic mean field (RMF) theory extended to the rotating 
frame has been applied in a systematic way to the description of 
SD bands, observed in the mass regions shown in Fig.\ \ref{exp-zoo}, 
and in the present contribution, these applications are briefly 
overviewed with stress on the description of the quantities 
extracted from $\gamma$ transitions within the bands and compared 
with some results obtained in non-relativistic mean field 
approaches. 

\section{Relativistic mean field theory in rotating 
frame}
\label{RMFrotframe}

  RMF theory describes the nucleus as a system of nucleons 
(Dirac spinors) which interact in a relativistic covariant 
manner through the exchange of virtual mesons \cite{SW.86}: 
the isoscalar scalar $\sigma$ meson responsible for the large 
scalar attraction at intermediate distances, the isoscalar 
vector $\omega$ meson responsible for the vector repulsion at 
short distances and the isovector vector $\rho$ meson which takes 
care for the asymmetry properties of nuclei with large neutron 
or proton  excess. In addition, the photon field $(A)$ accounts 
for the electromagnetic interaction. 

  RMF theory has been extended for the description of rotating 
nuclei (cranked relativistic mean field (CRMF) theory) 
by employing the concepts of the cranking model in Refs.\ 
\cite{KR.89,KR.90,KR.93} (see also Refs.\ \cite{KNM.93,MM.97} 
for alternative derivations). In the first applications 
pairing correlations were neglected and one-dimensional
cranking approximation was used. The CRMF equations include 
the Dirac equation in the Hartree-approximation for fermions
\begin{eqnarray}
\left\{\bff\alpha(-i\bff\nabla-\bff V(\bff r))~+~
V_0(\bff r)~+~\beta(m+S(\bff r))-{\sl\Omega}_x\hat{J}_x\right\}
\psi_i~=~\epsilon_i\psi_i
\label{DirCRMF}
\end{eqnarray}
where $V_{0}({\bf r})$ represents a repulsive vector potential, 
$S({\bf r})$ an attractive scalar potential, $\bff V(\bff r)$ 
the magnetic potential and the term $\Omega_x \hat{J_x}$ the Coriolis 
field. The time-independent inhomogeneous Klein-Gordon equations 
for the mesonic fields are given by
\begin{eqnarray}
\left\{-\Delta-({\sl\Omega}_x\hat{L}_x)^2+m_\sigma^2\right\}~
\sigma(\bff r)&=&
-g_\sigma\left[\rho_s^p(\bff r)+\rho_s^n(\bff r)\right]
-g_2\sigma^2(\bff r)-g_3\sigma^3(\bff r),
\nonumber \\
\left\{-\Delta-({\sl\Omega}_x\hat{L}_x)^2+m_\omega^2\right\}
\omega_0(\bff r)&=&
g_\omega\left[\rho_v^p(\bff r)+\rho_v^n(\bff r)\right],
\nonumber \\
\left\{-\Delta-({\sl\Omega}_x(\hat{L}_x+\hat{S}_x))^2+
m_\omega^2\right\}~
\bff\omega(\bff r)&=&
g_\omega\left[\bff j^p(\bff r)+\bff j^n(\bff r)\right],
\label{KGCRMF}
\end{eqnarray}
with source terms involving the various nucleonic densities
and currents
\begin{eqnarray}
\rho_s^{n,p}(\bff r) = \sum_{i=1}^{N,Z}(\psi_i(\bff r))^{\dagger} \hat{\beta} 
\psi_i(\bff r),\qquad & &
\rho_v^{n,p} (\bff r) = \sum_{i=1}^{N,Z} (\psi_i (\bff r))^{\dagger} \psi_i
(\bff r), \nonumber \\
\bff j^{n,p}(\bff r) &=&\sum_{i=1}^{N,Z} (\psi_i(\bff r))^{\dagger}
\hat{\bff\alpha} \psi_i (\bff r), 
\label{curr}
\end{eqnarray}
where the labels $n$ and $p$ are used for neutrons and protons, 
respectively.  In the equations above, the sums run over the 
occupied positive-energy shell model states only ({\it no-sea 
approximation}) \cite{RRM.86}. For simplicity, the equations for 
the $\rho$ meson and the Coulomb fields are omitted in Eqs.\ (\ref{KGCRMF}) 
since they have the structure similar to the equations for the 
$\omega$ meson. Since the coupling constant of the 
electromagnetic interaction is small compared with the coupling 
constants of the meson fields, the Coriolis term for the Coulomb 
potential $A_0(\bff r)$ and the spatial components of the vector 
potential $\bff A(\bff r)$ are neglected in the calculations. 

  CRMF theory has been used extensively in the study of SD bands 
in which the pairing correlations are expected to play only a minor 
role. This approach has an advantage since the question of a
self-consistent description of pairing correlations in finite 
nuclei starting from a relativistic Lagrangian still remains a not 
fully solved theoretical problem \cite{KurcR.91}. Considering, however,
that the pairing is a genuine non-relativistic effect, which 
plays a role only in the vicinity of the Fermi surface one can 
consider the pairing correlations only between the baryons  
using the phenomenological Gogny interaction with finite range
\begin{eqnarray}
V^{pp}(1,2) &=& \sum_{i=1,2} e^{-[({\bff r}_1-{\bff r} _2)/\mu_i]^2} 
\label{Gforce} \nonumber \\
& & \times (W_i+B_i P^{\sigma}- H_i P^{\tau} - M_i P^{\sigma} P^{\tau})
\end{eqnarray}
in the particle-particle (pairing) channel. In conjuction with 
RMF theory such an approach to the description of pairing correlations
has been applied, for example, in the study of ground state properties
\cite{LVR.98}, neutron halos \cite{PVLR.97}, and deformed proton 
emitters \cite{VLR.99}. The development of Cranked Relativistic 
Hartree-Bogoliubov (CRHB) 
theory using this concept for the description of pairing correlations has been 
finished recently and first results have been presented at 
the present conference. 

  The CRHB equations for the fermions in the rotating frame 
are given in one-dimensional cranking approximation by 
\begin{eqnarray}
\pmatrix{ h - \Omega_x \hat{J}_x    & \hat{\Delta}  \cr
-\!\hat{\Delta}^* &     -h^* + \Omega_x \hat{J}^*_x \cr}
\pmatrix{ U_k \cr V_k } =
E_k \pmatrix{ U_k \cr V_k }
\label{CRHB}
\end{eqnarray}
where $h=h_D-\lambda$ is the single-nucleon Dirac Hamiltonian 
minus the chemical potential $\lambda$ and $\hat{\Delta}$ is 
the pairing potential. $U_k$ and $V_k$ are quasiparticle Dirac 
spinors and $E_k$ denote the quasiparticle energies. The 
structure of time-independent inhomogeneous Klein-Gordon 
equations for the mesonic fields is the same as in Eqs.\ (\ref{KGCRMF}) 
with the exception that the probabilities of the occupation of different 
orbitals are taken into account, when nucleonic densities and currents
are calculated, by
\begin{eqnarray}
\rho_s^i(\bff r) = \sum_{k>0} 
 (V_k^i(\bff r))^{\dagger} \hat{\beta} V_k^i (\bff r), \qquad & &
\rho_v^i(\bff r) = \sum_{k>0} 
 (V_k^i(\bff r))^{\dagger} V_k^i (\bff r) 
\nonumber \\
\bff j^i(\bff r) &=&  \sum_{k>0} 
(V_k^i(\bff r))^{\dagger} \hat{\bff\alpha} V_k^i (\bff r).
\label{sourceHB}
\end{eqnarray}
  The sums over $k>0$ run over all quasiparticle states corresponding 
to positive energy single-particle states ({\it no-sea approximation})
and the indexis $i$ could be either $n$ (neutrons) or $p$ (protons). 
An additional feature is that in CRHB theory we go beyond the mean 
field and perform an approximate particle number projection before 
the variation by means of the Lipkin-Nogami method
\cite{L.60,N.64,PNL.73}. It turns out that this feature is 
extremely important for a proper description of the moments of 
inertia. 

  The spatial components of the vector mesons give origin to a magnetic 
potential $\bff V (\bff r)$ which breaks time-reversal symmetry and 
removes the degeneracy between nucleonic states related via this
symmetry \cite{KR.93,AKR.96}. This effect is commonly referred as  
{\it nuclear magnetism} \cite{KR.89}. It is very important for a proper 
description of the moments of inertia \cite{KR.93}. Consequently, 
the spatial components of the vector $\omega$ and $\rho$ mesons are 
properly taken into account in a fully self-consistent way in the 
calculations. 

 The microscopic relativistic and non-relativistic mean field
approaches being fully self-consistent share some common features, 
for example, such as the low effective mass being typically in the 
range of $0.6-0.7$ and
the presence of time-odd mean fields which have a large impact
on the moments of inertia. The low values of the effective mass 
in microscopic theories leads to a low level density in the vicinity 
of the Fermi level compared with experiment. This problem can, in 
general, be cured by taking into account the coupling between 
single-particle motion and low-lying collective vibrations, see 
Ref.\ \cite{QF.78}. The time-odd mean fields, which play an 
important role in rotating nuclei, are defined uniquely in the 
RMF theory, while in the approaches based on effective interaction 
of Skyrme type this depends on underlying energy functional or
the corresponding effective two-body interaction \cite{DD.95}. 
A clear advantage of RMF theory is the fact that the spin-orbit 
term emerges in a natural way \cite{R.96}, while it has to be 
parametrized in the non-relativistic approaches \cite{RS.80}. 
RMF theory is distinct from non-relativistic theories in the 
mechanism of nuclear saturation \cite{R.96}, where the nucleonic 
potential emerges as a difference of large attractive scalar 
($S({\bf r})$) and repulsive vector ($V_{0}({\bf r})$) potentials.

Theoretical approaches, based on the non-relativistic macroscopic + 
microscopic method, which use the Woods-Saxon 
\cite{NWJ.89,WCNWJ.91,NDBBR.85} or Nilsson \cite{BR.85}
potentials for the microscopic part still remain very 
powerful tools for our understanding of rotating nuclei. By 
treating the bulk and the single-particle properties separately 
they have the advantage to fit many details of the actual nuclei  
directly to the appropriate region under investigation. However, 
this separation leaves some room for inconsistencies between 
the macroscopic and microscopic parts as illustrated, for example, 
in Refs.\ \cite{Dud84,KRA.98}. In addition, time-odd mean 
fields are neglected in this method. However, 
these phenomenological models have some advantages related
to the facts that (i) the effective mass is one by definition,
(ii) they are more flexible due to the separation of bulk and
single-particle properties, (iii) the numerical calculations 
are by orders of magnitude less time consuming than the 
ones in the microscopic approaches (see, 
for example, the discussion in Ref.\ \cite{PhysRep}).

\section{Superdeformation in the regime of weak pairing 
correlations.}

 CRMF theory has been extensively used for the description 
of SD bands in which the pairing correlations are expected to 
be considerably quenched. Detailed investigations have been 
performed using this approach in the $A\sim 140-150$ and in 
the $A\sim 60$ mass regions. One should clearly recognize that 
the neglect of pairing correlations is an approximation because
pairing correlations being weak are still present even at the highest 
rotational frequencies. However, the question of the description 
of pairing correlations in the regime of weak pairing still 
remains an open problem (see, for example, the introduction in Ref. \
\cite{AKR.96}). Almost all cranked mean field approaches which 
aim to describe pairing correlations in rotating nuclei 
use an approximate particle number projection before variation 
by means of the Lipkin-Nogami method, see for example Ref.\ \cite{SWM.94}.
However, the applicability of this method in the regime of weak
pairing correlations as an approximation to exact particle number 
projection seems questionable \cite{ZSF.92,DN.93,Mag93}. Keeping 
this in mind we think that the neglect of pairing correlations is 
a reasonable approximation in a specific physical situations and, 
although some specific features such as, for example, paired band 
crossings cannot be addressed, it allows to gain considerable 
understanding of physical phenomena in the high spin region.

\subsection{The $A\sim 140-150$ mass region}

  Since the discovery of superdeformation in $^{152}$Dy in 1986 
\cite{TNN.86}, this mass region was the testing field for different 
theoretical models and concepts. Already the investigations within 
the macroscopic + microscopic method gave an understanding of the 
impact of different single-particle orbitals on the dynamic moments 
of inertia $J^{(2)}$ and transition quadrupole moments $Q_t$ 
\cite{BRA.88,NWJ.89}, possible underlying mechanisms of identical 
bands (see Ref.\ \cite{BHN.95} and references therein), methods 
of configuration assignment based on properties of dynamic moments 
of inertia \cite{BRA.88} or effective alignments $i_{eff}$ 
\cite{Rag91,Rag93} (or similar methods), considerable quenching of 
pairing correlations \cite{NWJ.89,SGB.89} and so on. These
developments are covered in review articles \cite{AFN.90,BHN.95}. 

  A systematic investigation of the properties of the SD bands in
this mass region has been performed in the framework of the CRMF 
theory in Refs.\ \cite{AKR.96,ALR.98,Hung,RAM.97,RA.97,CRMF-ho153}. 
Using mainly the NL1 parametrization of RMF Lagrangian \cite{RRM.86},
which has only 7 parameters fitted in the middle of 80ties to the 
properties of several spherical nuclei, it was shown that  
CRMF theory reproduces well the experimentally observed features 
such as the properties of dynamic moments of inertia $J^{(2)}$, 
single-particle ordering in the SD minimum, alignment properties 
of single-particle orbitals etc. Calculated charge quadrupole 
moments $Q_0$ are in general somewhat larger than the average 
experimental values quoted in literature but they are still within 
the experimental error bars if the uncertainties due to stopping 
powers ($\sim 15\%$) are taken into account. The last feature is 
also typical for other theoretical approaches based either on 
the Nilsson or the Woods-Saxon potentials \cite{NWJ.89,Rag93} 
and Skyrme forces \cite{SDDN.96} (see also the discussion in 
Ref.\ \cite{AKR.96}). 

\begin{figure}
\epsfxsize 12.0cm
\epsfbox{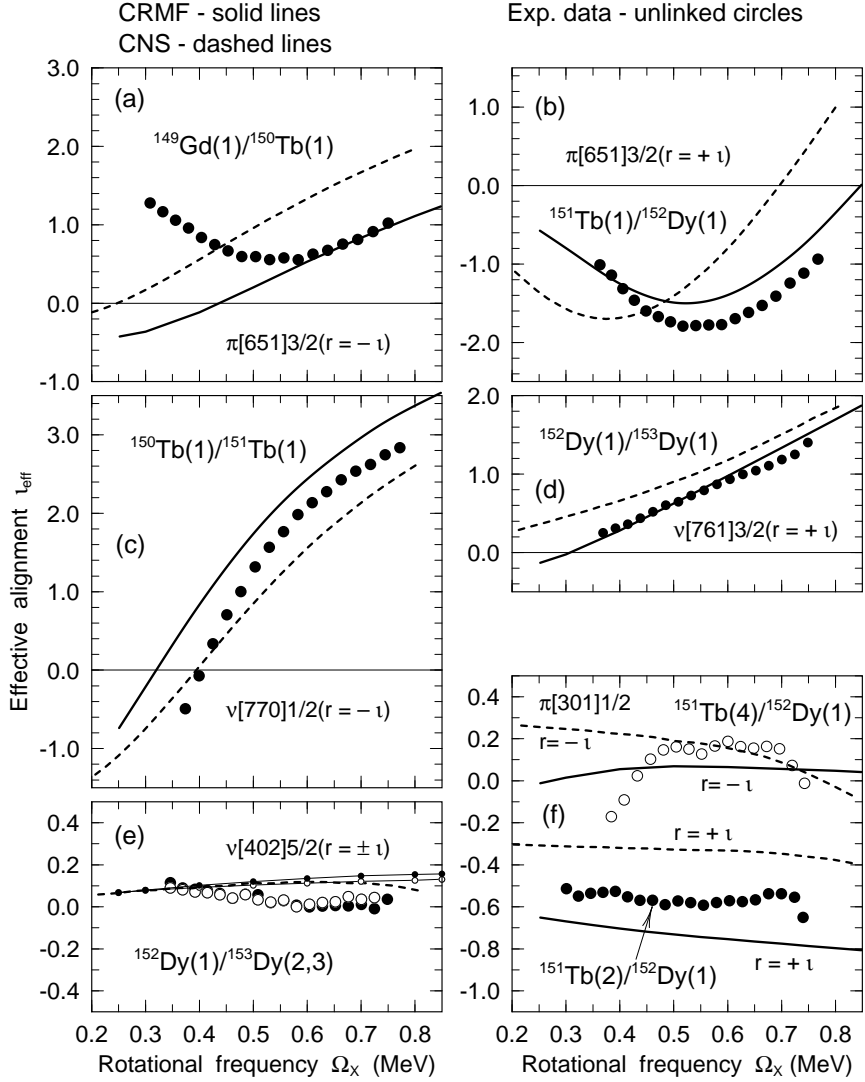}
\caption{Effective alignments, $i_{eff}$ (in units $\hbar$),
extracted from experimental data are compared with those 
extracted from the corresponding configurations calculated
in the CRMF theory and in the configuration-dependent cranked
Nilsson-Strutinsky (CNS) approach based on the Nilsson potential. 
The experimental effective alignment between bands A and B is 
indicated as `A/B'. The band A in the lighter nucleus is taken as
a reference, so the effective alignment measures the effect
of the additional particle. The compared calculated configurations 
differ in the occupation of the orbitals shown in the panels.
(From results obtained in Refs.\ \protect\cite{ALR.98,Rag93}).}
\label{fig-align}
\end{figure}

  The overview of all these results definetely goes beyond the
scope of the present article, especially considering the fact
that Refs. \cite{ALR.98,AKR.96,Hung,CRMF-ho153} contain detailed 
discussions of the relation between CRMF results and the ones 
obtained in non-relativistic mean field approaches. In the 
present review we would like to point out that the CRMF theory 
is able to describe reasonably well the alignment properties of 
the single-particle orbitals despite the fact that no information
on single-particle properties has been taken into account during 
the fit of the parameters of the RMF theory. Fig.\ \ref{fig-align} 
compares the effective alignments of specific single-particle 
orbitals active at superdeformation in the nuclei around $^{152}$Dy 
obtained in the CRMF theory and cranked Nilsson-Strutinsky (CNS) 
approach based on the Nilsson potential with experimental values. 
The CRMF theory reproduces in average the experimental $i_{eff}$ 
values better than the CNS approach. The descripancy between 
CRMF calculations and experiment seen in Fig.\ \ref{fig-align}a at 
$\Omega_x \leq 0.5$ MeV is most likely due to the fact that 
pairing correlations play some role at low rotational frequency 
in the $^{149}$Gd(1) band, see Ref.\ \cite{ALR.98}. One should 
note, however, that with exception of the $[651]3/2(r=\pm i)$ 
orbitals the results of the CNS calculations are also reasonably 
close to experimental data. 

Effective (or relative) alignments are important ingredients of our 
understanding of the underlying mechanism of the appearance of identical 
and non-identical bands \cite{BHN.95}. Indeed, the fractional change 
in $J^{(2)}$ of two bands A and B (the criterion most frequently used 
for the selection of identical bands) is proportional to the derivative 
$di_{eff}^{A,B}/d\Omega_x$ \cite{BHN.95}. Since the slope of 
experimental and theoretical $i_{eff}$ values as a function of 
$\Omega_x$ is similar in many cases (see Fig.\ \ref{fig-align}), it 
is clear that also the difference between $J^{(2)}$ values of compared
bands is reproduced reasonably well in these cases (see Ref.\ 
\cite{ALR.98}). Systematic investigation of effective (relative)
alignments in this mass region has been performed so far only
in the CNS approach based on the Nilsson potential (see Refs.\ 
\cite{Rag91,Rag93,Lennart} and references quoted in Ref.\ \cite{ALR.98})
and in the CRMF theory \cite{ALR.98,Hung,CRMF-ho153}. So far, only
the effective alignments associated with the $\pi [301]1/2$ orbital
have been studied within the cranked Hartree-Fock approach using
variuos Skyrme forces \cite{DD.95}.

   Phenomenological approaches based on cranked Nilsson
or Woods-Saxon potentials have been extensively used in this mass 
region, see Ref.\ \cite{D.99} and references therein. Among the 
microscopic non-relativistic approaches only the approaches 
(CHF or cranked Hartree-Fock-Bogoliubov (CHFB)) 
based on the effective forces of the Skyrme type have been used 
in a systematic way in nuclei around $^{152}$Dy 
\cite{DD.95,BFH.96,EDD.97,RBFH.99}. While in the CRMF theory,
the calculated dynamic moments of inertia $J^{(2)}$ are close
to experimental data (see Refs. \cite{AKR.96,Hung,CRMF-ho153}),
the calculations based on Skyrme forces overestimate the 
magnitude of $J^{(2)}$ typically by 10\%.

\subsection{The $A\sim 60-70$ mass region}

\begin{figure}
\epsfxsize 16.0cm
\epsfbox{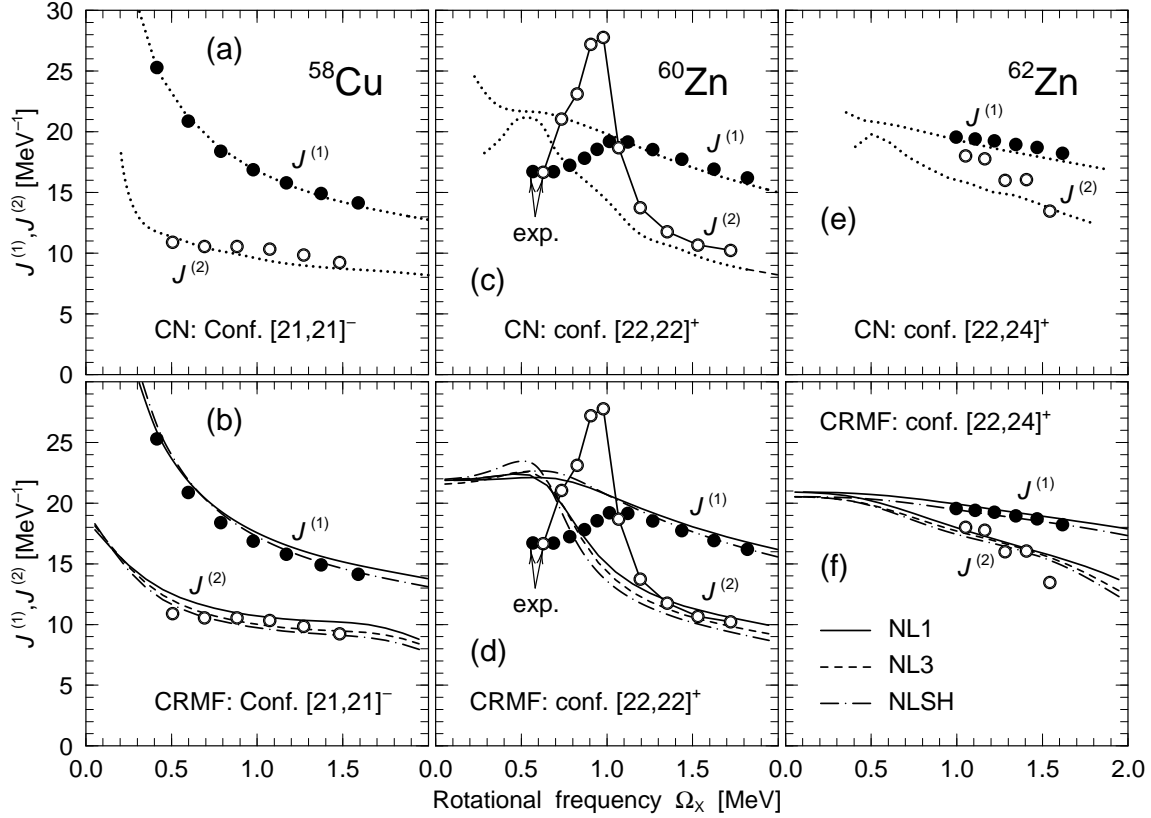}
\caption{Kinematic $J^{(1)}$ (unlinked solid circles)
and dynamic $J^{(2)}$ (open circles) moments of inertia
of observed bands versus the ones of assigned calculated
configurations. The notation of the lines is given in the
figure. To label the configurations we use the shorthand 
notation $[p_1p_2,n_1n_2]$ where $p_1$ $(n_1)$ is the number 
of proton (neutron) $f_{7/2}$ holes and $p_2$ $(n_2)$ is the 
number of proton (neutron) $g_{9/2}$ particles. Superscripts 
to the configuration labels (e.g. $[22,22]^+$) are used to 
indicate the sign of the signature $r$ for that configuration 
$(r=\pm 1)$. The values of $J^{(1)}$ calculated with NL3 
are typically in between the ones obtained with NL1 and NLSH,
so for simplicity they are not shown. (From 
Ref.\ \protect\cite{A60}).}
\label{fig-a60}
\end{figure}

 As illustrated in Fig.\ \ref{exp-zoo}, the rotational 
properties of the SD band in $^{60}$Zn are distinctly 
different from the ones seen in the $A\sim 150$ and 190 
mass regions. Indeed, at the highest rotational frequencies 
the dynamic moment of inertia $J^{(2)}$ is only approximately 
60\% of the kinematic moment of inertia $J^{(1)}$. The 
comparative study of SD and highly deformed bands in this 
mass region has been performed within the framework of the 
CRMF theory and CNS approach 
based on the Nilsson potential in Refs.\ \cite{A60,Zn60,Zn68}.
Defining the spin values of the unlinked SD and highly deformed
bands by means of effective alignment approach \cite{Rag91}, it was shown
that also other bands of these types show the same relation 
between $J^{(1)}$ and $J^{(2)}$, see Fig.\ 
\ref{fig-a60}. These features are very similar to the ones seen 
in smooth unfavoured terminating bands observed in the $A\sim 110$ 
mass region \cite{Rag95,A110,PhysRep} and in a similar way they have 
been attributed to the influence of the limited angular 
momentum content of the single-particle configurations of 
the observed bands. For example, in the case of the $^{60}$Zn 
SD band, one obtains from the distribution of particles and holes over
high- and low-$j$ orbitals the ``maximum'' spin 
of the $[22,22]^+$ configuration as $I=36^+$. As a result, this 
band is three transitions away from the ``maximum'' spin. However, 
the states of the ``maximum'' spin are calculated triaxial both in 
the CRMF and in the CNS approaches. Thus, contrary to the smooth 
terminating bands, the SD bands do not terminate in the non-collective
single-particle state at ``maximum'' spin in the calculations. This 
behavior can be understood as caused by the interaction between 
low-$j$ and high-$j$ orbitals in the $N=3$ shell at high rotational 
frequencies \cite{A60}. The calculations suggest that these bands can 
be continued beyond the ``maximum'' spin in a similar way as it 
happens in the cranked harmonic oscillator, see Ref.\ \cite{PhysRep} 
and references therein. 

 The above mentioned general features of the SD bands in the 
$A\sim 60-70$ mass region are in line with a previous study 
\cite{Rag87}, where it was shown that the relation between 
$J^{(1)}$ and $J^{(2)}$ is not so much 
determined by the deformation (at $I=0$), but rather how far away 
the band is from its ``maximum'' spin value. For example, large 
differences between $J^{(1)}$ and $J^{(2)}$ are not expected in 
the SD bands in the $A\sim 150$ mass region (see middle panel in 
Fig.\ \ref{exp-zoo}) because the ``maximum'' spin of their 
configurations, $I_{max}\sim (150-200)\hbar$, is far above the 
experimentally accessible spin values (see Ref.\ \cite{Rag87}).
An additional consequence of the fact that SD and highly deformed 
bands in the mass region of interest are coming close to the 
``maximum'' spin is a gradual drop of collectivity (i.e. a drop
of the transition quadrupole moment $Q_t$) with increasing spin 
which is larger than the one expected in the $A\sim 150$ mass 
region, see Fig. 4 in Ref.\ \cite{A60} and Ref.\ \cite{AKR.96}.

 It turns out that the rotational properties ($J^{(1)}$ and 
$J^{(2)})$\footnote{The description of the paired band crossing
observed in $^{60}$Zn band at $\Omega_x \sim 1.0$ MeV is not 
addressed in the present calculations without pairing.}, 
effective alignments $i_{eff}$ and transition 
quadrupole moments $Q_t$ of highly deformed and SD bands 
are described very well  both in the CRMF and in the CNS 
approaches, see Refs.\ \cite{A60,Zn60,Zn68} and Fig.\ \ref{fig-a60} 
in the present manuscript, with the results of CRMF calculations 
being in average in better agreement with experimental data.   

 The structure of the highly deformed and SD bands in this mass 
region has been studied also in other approaches. The 
CHF calculations with an effective forces of the Skyrme type have 
been performed in Refs.\ \cite{Cu58,Ni56,Rud98}. The SD band 
observed in $^{62}$Zn has been extensively studied both in the 
CRMF theory and in the CHF approach with Skyrme forces in 
Ref.\ \cite{MM.99} using different parametrizations of the
RMF Lagrangian and effective forces. The projected shell model
which goes beyond mean field approximation has been used in
the study of SD bands in Ref.\ \cite{PSMa60}.


\section{Superdeformation in the $A\sim 190$ mass 
region.}
\label{SDweakpair}
   
\begin{figure}
\epsfxsize 12.0cm
\epsfbox{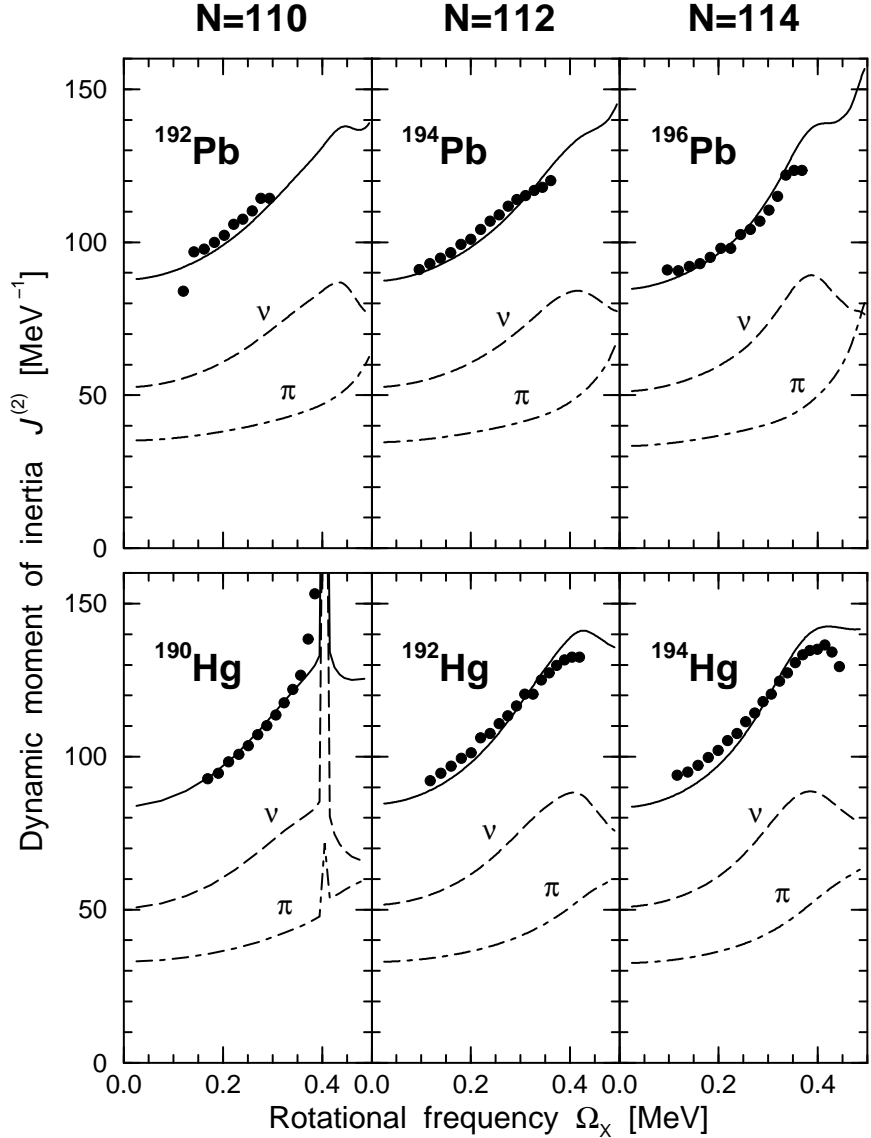}
\caption{Dynamic moments of inertia $J^{(2)}$ of observed 
(solid circles) yrast SD bands in the $^{190,192,194}$Hg and
$^{192,194,196}$Pb nuclei versus the ones of calculated lowest 
in energy SD configurations. Solid lines show the total calculated 
dynamic moments of inertia $J^{(2)}$, while long-dashed and 
dash-dotted lines show the contribution in $J^{(2)}$ from neutron 
and proton subsystems. The experimental data are taken from 
Refs.\ \protect\cite{Pb192a} ($^{192}$Pb), 
\protect\cite{Pb194a,Pb194b,Pb194c} ($^{194}$Pb), 
\protect\cite{Pb196a} ($^{196}$Pb), \protect\cite{Hg190} 
($^{190}$Hg), \protect\cite{Hg192} ($^{192}$Hg) and 
\protect\cite{Hg194a,Hg194b} ($^{194}$Hg).}
\label{sys-j2}
\end{figure}

\begin{figure}
\epsfxsize 12.0cm
\epsfbox{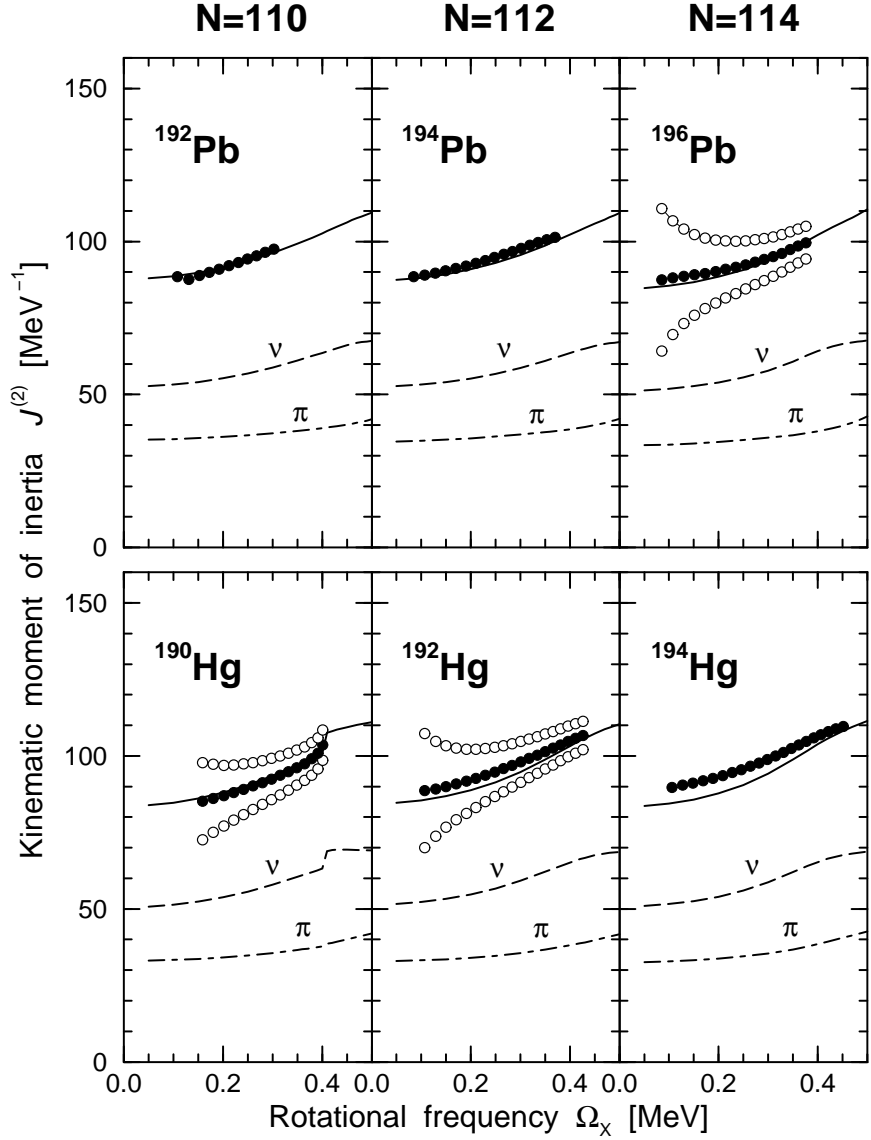}
\caption{The same as Fig.\ \protect\ref{sys-j2} but for kinematic 
moments of inertia $J^{(1)}$. Only in the cases of the yrast SD 
bands in $^{192,194}$Pb \protect\cite{Pb192a,Pb194b,Pb194c}
and $^{194}$Hg \protect\cite{Hg194b}, the spins of the bands are 
firmly ($^{194}$Hg, $^{194}$Pb) or tentatively ($^{192}$Pb) 
defined in experiment. In other cases, the 'experimental' kinematic 
moments of inertia $J^{(1)}$ are shown for three different spin 
values of the lowest state $I_0$ in SD band, which are consistent 
with the signature of the lowest calculated SD configuration. 
The $J^{(1)}$ values obtained in such a way 
are shown by circles with values being in best agreement with 
calculations indicated by solid circles. These comparisons 
indicate that the lowest transitions in the yrast SD bands of 
$^{190}$Hg, $^{192}$Hg and $^{196}$Pb with energies 316.9, 214.4 
and 171.5 keV, respectively, most likely correspond to the spin 
changes of $14^+ \rightarrow 12^+$, $10^+ \rightarrow 8^+$ and 
$8^+ \rightarrow 6^+$.}
\label{sys-j1}
\end{figure}

 The SD bands in this mass region are in most cases characterised 
by the dynamic moment of inertia $J^{(2)}$ which increases with 
increasing rotational frequency, see Fig.\ \ref{sys-j2}. 
Kinematic moments of inertia $J^{(1)}$ of linked SD 
bands in $^{192,194}$Pb and $^{194}$Hg (see Fig.\ \ref{sys-j1}) 
show the same features and in addition the relation 
$J^{(2)}\geq J^{(1)}$ holds. This indicates that the pairing 
correlations play a more important role in these bands compared 
with the ones observed in the $A\sim 60$ and 150 mass regions, 
see Sect.\ \ref{SDweakpair}.

 Different non-relativistic approaches have been employed for the 
study of the superdeformation in this mass region. The CHFB 
calculations of SD 
bands in this mass region have been performed using Gogny forces 
without \cite{GDBL.94} and with \cite{GDBPL.97,VER.97,VE.97} 
approximate particle number projection before variation by means 
of Lipkin-Nogami method. The calculations using Gogny forces with 
diagonalization of the collective Bohr Hamiltonian with the 
inertial functions calculated at zero spin via the Gaussian 
Overlap Approximation to the Generator Coordinate Method have 
been performed in Refs.\ \cite{GDBPL.97,LBDG.93}. A clear advantage 
of the microscopic theories based on the effective forces of the 
Gogny type is that no additional parameters are needed to describe 
the pairing correlations.

 On the other hand, theoretical approaches discussed below use
some assumptions about the type of pairing interaction and fit
its parameters to the experimental data. An extensive investigation 
of SD bands in this mass region has been performed within the CHFB 
approach with effective forces of the Skyrme type in Refs.\
\cite{GBDFH.94,HBF.95,THBDF.95,TFHB.97,HJ.98}. It was concluded  
that special attention should be taken to the effective interaction 
in the pairing channel  and that surface active delta pairing gives 
a better agreement with data for the behaviour of the dynamic moment 
of inertia $J^{(2)}$ versus rotational frequency than either the 
volume active or the seniority pairing \cite{THBDF.95}. The CHFB 
approach based on the Woods-Saxon potential together with
Lipkin-Nogami method for approximate particle number projection 
has been developed in Refs.\ \cite{WS.94,WS.95}. This approach has 
been used frequently in the investigation of SD bands observed in 
different mass regions (see Ref.\ \cite{D.99} and references quoted 
therein). It employs monopole and quadrupole pairing in the 
particle-particle (pairing) channel.  Monopole+quadrupole pairing 
has also been used in the projected shell model calculations of the 
SD bands in this mass region \cite{PSM}. Note, however, that the
later approach goes beyond the mean field approximation.
 
 Cranked Relativistic Hartree-Bogoliubov theory (see
Sect. \ref{RMFrotframe}) has been applied to the 
description of the SD bands observed in even-even nuclei with 
neutron numbers $N=110,112,114$ \cite{AKR.99}. The calculations 
have been performed with the NL1 parametrization \cite{RRM.86} 
of the relativistic mean field Lagrangian and the D1S set of the 
parameters for the Gogny force \cite{D1S}. The results of the 
calculations for the dynamic and the kinematic moments of inertia 
are shown in Figs.\ \ref{sys-j2}, \ref{sys-j1}. One can see that 
a very successful description of rotational features 
of experimental bands is obtained in the calculations without 
adjustable parameters. A comparison with results of other 
approaches based on the mean field concept (see, for example, 
Fig. 3 in Ref. \cite{WS.94}, Figs.\ 1 and 8 in Ref.\
\cite{THBDF.95}, Fig.\ 1 in Ref.\ \cite{VER.97}, Figs.\ 5 and 6 in 
Ref.\ \cite{VE.97} and Fig.\ 1a in Ref. \cite{GDBPL.97}) 
and the projected shell model (see Figs.\ 1 and 2 in Ref.\ \cite{PSM})
indicates that the CRHB calculations provide one of best agreements 
with experimental data. The calculated values of transition quadrupole 
moments $Q_t$ are also close to the measured ones, however, more 
accurate and consistent experimental data on $Q_t$ is needed 
in order to make detailed comparisons between experiment and 
theory, for details see the discussion in Ref.\ \cite{AKR.99}.
RMF theory also excellently reproduces the excitation energies 
of the SD bands relative to the ground state in $^{194}$Hg and
$^{194}$Pb nuclei \cite{LR.98}.

 The increase of kinematic and dynamic moments of inertia in this 
mass region can be understood in the framework of CRHB theory as 
emerging predominantly from a combination of three effects: the 
gradual alignment of a pair of $j_{15/2}$ neutrons, the alignment 
of a pair of $i_{13/2}$ protons at a somewhat higher frequency, 
and decreasing pairing correlations with increasing rotational 
frequency. The interplay of alignments of neutron and proton
pairs is more clearly seen in the Pb isotopes where the calculated
$J^{(2)}$ values show either a small peak (for example, 
at $\Omega_x \sim 0.45$ MeV in $^{192}$Pb, see Fig.\ \ref{sys-j2}) 
or a  plateau (at $\Omega_x \sim 0.4$ MeV in $^{196}$Pb, 
see Fig.\ \ref{sys-j2}). With increasing rotational frequency, 
the $J^{(2)}$ values determined by the alignment in the neutron 
subsystem decrease but this process is compensated by the increase 
of $J^{(2)}$ due to the alignment of the $i_{13/2}$ proton pair. 
This leads to the increase of the total $J^{(2)}$-value at 
$\Omega_x \geq 0.45$ MeV. The shape of the peak (plateau) in $J^{(2)}$
is determined by a delicate balance between alignments in the proton 
and neutron subsystems which depends on deformation, rotational 
frequency and Fermi energy. For example, no increase in the total 
dynamic moment of inertia $J^{(2)}$ has been found in the calculations
after the peak up to $\Omega_x=0.5$ MeV in $^{192}$Hg, see 
Fig.\ \ref{sys-j2}. It is also of interest to mention that the sharp 
increase in $J^{(2)}$ of the yrast SD band in $^{190}$Hg is also 
reproduced in the present calculations. One should note that the 
CRHB calculations slightly overestimate the magnitude of $J^{(2)}$ 
at the highest observed frequencies. The possible reasons could be 
the deficiencies either of the Lipkin-Nogami method \cite{Mag93} or 
the cranking model in the band crossing region or both of them. 

\section{Conclusions}

 The applications of the cranked versions of the relativistic mean 
field theory to the description of superdeformed bands in different 
mass regions clearly indicates that this theory provides very good 
agreement with experimental data. The level of agreement is comparable
and in many cases better than the one obtained in the non-relativistic
approaches.

  We would like to express our gratitude to our collaborators, who 
contributed to the investigations presented here, namely, to 
J.\ K{\"o}nig, G.\ A.\ Lalazissis and I.\ Ragnarsson. A.\ V.\ A. 
acknowledges support from the Alexander von Humboldt Foundation. 
This work is also supported in part by the Bundesministerium 
f{\"u}r Bildung und Forschung under the project 06 TM 875.



\newpage

\end{document}